\documentstyle[12pt]{article}

\def\lax{{$\mathrel{\hbox{\rlap{\hbox{\lower4pt\hbox{$\sim$}}}\hbox{$<$}}}$}}
\def\gax{{$\mathrel{\hbox{\rlap{\hbox{\lower4pt\hbox{$\sim$}}}\hbox{$>$}}}$}}
\def\simlt{\lower.5ex\hbox{$\; \buildrel < \over \sim \;$}}
\def\simgt{\lower.5ex\hbox{$\; \buildrel > \over \sim \;$}}

\def\percm2{cm$^{-2}$}

\begin{document}

\centerline{\Large{\underbar {\bf The Carnegie Observatories Astrophysics Series}}}

\vspace*{+0.1cm}
\centerline{\it Luis C. Ho, Series Editor}

\vspace*{+1.0cm}

\noindent
On the occasion of the Centennial of the Carnegie Institution of Washington,
Carnegie Observatories held a series of four astrophysics symposia in Pasadena
from October 2002 to February 2003.  The topics of the symposia were: 
\vspace*{+0.3cm}

\begin{enumerate}

\item
{\it Coevolution of Black Holes and Galaxies} [Ed. L. C. Ho]

\item
{\it Measuring and Modeling the Universe} [Ed. W. L. Freedman] 

\item
{\it Clusters of Galaxies: Probes of Cosmological Structure and Galaxy 
    Evolution} [Ed. J. S. Mulchaey, A. Dressler, and A. Oemler] 

\item
{\it Origin and Evolution of the Elements} [Ed. A. McWilliam and M. Rauch] 
\end{enumerate}

\vspace*{+0.3cm}

\noindent
The invited papers of the symposia, which have been peer-reviewed and
carefully edited, will be published in 2004 by Cambridge University Press, as 
the first four volumes of the {\it Carnegie Observatories Astrophysics 
Series}.  The papers from the contributed talks and posters, along with the 
full set of the invited papers, are available electronically at 
\vspace*{+0.3cm}

\centerline{{\tt http://www.ociw.edu/ociw/symposia/series/}  }

\vspace*{+0.3cm}
\noindent
The purpose of this note is to 
alert the community of the availability of this resource.

\end{document}